\documentstyle[editedbook,epsfig,psfig,epsf]{mq}

\def\cm2{cm$^2$ }
\def\se1{s$^{-1}$ }

\begin{opening}
\title{Microquasars in the low/hard state: strong coronae, compact
jets, and the high frequency variability}
\author{A. Merloni$^1$}
\institute{$^1$ Max-Planck Institut f\"ur Astrophysik,
Karl-Schwarzschild-Str. 1, D-85740 Garching bei M\"unchen, Germany}
\end{opening}

\runningtitle{Microquasars in the low/hard state}
\runningauthor{Merloni}

\begin{document}
\vspace{-0.5cm}
\begin{abstract}
{\small We apply a model of magnetically dominated coronae above
  standard accretion discs to the low/hard state of galactic black
  holes. When the disc-corona coupling is accounted for
  self-consistently assuming that magneto-rotational instability is at work
  in the disc, and that the corona is generated by buoyant escape of
  disc magnetic structures, then the model predicts powerful, X-ray
  emitting coronae at low accretion rates. A main consequence is
  discussed: the possibility that the corona itself
  is the launching site of powerful, MHD driven jets/outflows. This
  depends crucially of the coronal scaleheight. Finally, we present the
  first radial profiles of a corona a different accretion rates, and
  discuss their implications for high frequency variability.
}
\end{abstract}

\section{Introduction}
In \cite{mf02} we have shown how it is possible to build
a physically self-consistent model for an accretion disc corona: 
assuming that the 
turbulent magnetic stresses generated by MRI are responsible for angular
momentum transport in the disc, and that the field saturates mainly
due to buoyancy (as should be the case if strong coronae are to be
generated), then the fraction of power released into the
corona as a function of radius, $f(R)$, can be uniquely determined by
solving the algebraic equations for accretion disc structure.
For {\it uniform} discs (which may not be the case in the radiation
pressure dominated parts, see \cite{tss02}), we have
 \begin{equation}
\label{eq_fr}
f(R) \simeq \sqrt{\alpha} \left(1+\frac{P_{\rm
      rad}(R)}{P_{\rm gas}(R)}\right)^{-1/4},
\end{equation}
where $\alpha$ is a constant of the order of unity 
and $P_{\rm rad}$ and $P_{\rm gas}$ represent radiation and gas
pressure in the disc, respectively.

A major consequence of such a disc-corona coupling, which is of
interest in the context of microquasars, is that coronae are more
powerful at low accretion rates, in particular below the critical rate
at which the radiation pressure dominated part of the disc disappears
altogether. Indeed, if the $\alpha$ viscosity parameter is high
enough, $f$ can approach unity, and we are therefore left with a flow
in which most of the energy goes into the magnetically dominated
corona. The cold, geometrically thin disc, 
although very dim, may manifest itself as a
reprocessor of X-ray coronal radiation. This makes our
solution different from a simple magnetically dominated non radiative
accretion flow (NRAF): the two may be distinguished, for example, by
the presence of relativistically smeared reflection features.

\section{MHD outflows from powerful coronae}
A magnetically dominated
corona, with magnetic flux tubes corotating with the underlying disc
at nearly Keplerian speed, $v_{\rm Kep}$, is prone to generate powerful MHD
outflows. Let us examine in some detail the disc-corona-outflow
energetics.
At a distance $R$ from the center, the (magnetic) 
energy flux emerging from the disc is
\begin{equation}
Q_{\rm c}(R)=f(R)Q(R)=\frac{3 G M \dot M f(R)}{8 \pi R^3}\; 
{\rm ergs}\; {\rm cm}^{-2} \; {\rm s}^{-1},
\end{equation}
Part of this energy will be dissipated in reconnection events, 
heating the coronal plasma that is then cooled by inverse
Compton scattering soft photons. Such energy flux will
emerge as hard X-rays, and can be parametrized as
$F_{\rm H}(R)=(1-\eta)Q_{\rm c}(R)$, where $\eta$ is in general 
a function of the distance. The remaining fraction of the
coronal magnetic energy flux is carried off by the jet/outflow: 
$Q_{\rm j}(R)=\eta Q_{\rm c}(R)$.
Most models of MHD generated outflows from rotating
systems \cite{bz77,lop99} agree on the conclusion that the final power carried by such
jets depends on the {\it poloidal} component of the magnetic field and on the
angular velocity of the field lines. We have therefore
$dL_{\rm j}=\xi 2 \pi R Q_{\rm j}(R) dR=(B_{\rm p}^2/8 \pi) d{\cal A}_c
\Omega R$,
where 
$d{\cal A}_c \equiv \xi 2 \pi R dR$ is the element of disc area covered
by the corona (which has a covering fraction $\xi$) and the poloidal
component of the coronal magnetic field  is  a function  of the corona
scaleheight, $H_c$ (the height of a
reconnection site, from which the outflow is launched): 
$B_{\rm p}\simeq B (H_c/R)$.
The intensity of the coronal magnetic field, in turn, depends on how
quickly the field dissipates
$B^2=\frac{8 \pi Q_{\rm c}(1-\eta)}{\xi v_{\rm diss}}$.

Combining the expressions for $dL_{\rm j}$ and $B$, we obtain
\begin{equation}
\label{eq:eta}
\eta=\left(1+\xi \left(\frac{R}{H_c}\right)^2 \frac{v_{\rm
          Kep}(R)}{v_{\rm diss}}\right)^{-1}.
\end{equation}
The fraction of coronal power that is channeled in a MHD jet/outflow,
depends crucially on the height of a reconnection site (or,
more generally, on the scaleheight of the magnetically dominated
corona). 

\section{Radial profiles}
Most of the variability of black hole binaries 
(both the broad band one and the QPOs)
can be spectrally associated to the hard component
\cite{cgr01,rem01}. 
Moreover,
QPO frequencies tend to correlate with various X-ray spectral
parameters; in particular with those sensitive to the system size. In
general, higher frequencies are associated to softer spectra. In
principle,
if a magnetic corona is responsible for generating (or amplifying) the
observed  modulation of the hard X-ray flux, and the corona is
vertically extended (large $H_c$), both radial and vertical
oscillation may be observed. This implies that both the radial and
vertical size of the corona play an important role in shaping the
Power Density Spectra of these sources \cite{no02}.  
For example, if jet and corona are coupled systems, as we envisage, QPO
properties (in particular their amplitude and narrowness) are bound to
be influenced by the jet. Indeed, \cite{tru01} shows that when the
microquasar
GRS 1915+105 is in a hard state characterized by strong radio emission
(states $\chi_1$ and $\chi_3$ according to the classification of
\cite{bel00}), the high frequency noise component in the PDS is
strongly suppressed as compared to the hard states that do not show
radio emission (states $\chi_2$). 

Radial profiles of coronal emissivity and of the fraction of the total
gravitational power channeled either into coronal heating or
into an outflow, should be derived in order to build more physical
models for the QPO mechanism. As a final goal, time dependent
versions of such models should be implemented.
Here we present a simple, illustrative example of a possible
calculation. We have fixed the free parameters 
$\xi=0.5$ and $c/v_{\rm diss}=30$. 
The coronal optical depth is fixed at $\tau=0.8$ (variation of this
parameters affect the slope of the X-ray spectrum but not the overall energetics), and the
temperature is calculated self-consistently from the heating-cooling
balance (including both coronal synchrotron and
disc radiation fields as sources of soft photons for
Comptonization) as in \cite{mf01b}, at every annulus of width $dR$.
The most critical assumption is that regarding the radial
profile of the coronal scaleheight $H_c$, for the influence it has 
in determining the fraction of power
that goes into the jet. In general, the coronal scaleheight will be
larger for a larger value of $f$ (because more strongly buoyant flux
tubes should rise more) and proportional to the number
of twists a coronal loop, placed at a distance $R$ and 
 with footpoints separated by a distance $dR$, experience due to the
 disc differential rotation that torques the flux tube. Also, it should have a
power-law dependence on the distance due to the natural disc
flaring. In summary, we may write
$H_c/R \propto R^p f(R) d \ln \Omega/dR = A f(R) R^{p-1}$, with
$0<p<1$. In our calculations, we chose $p=1/2$ and $A=1$.

Once the accretion rate is fixed, we can integrate the disc
structure equations with the prescription of Eq. (\ref{eq_fr}) for the
coronal fraction, and solve for $f(R)$ and $\eta(R)$. With the
solution in hand, we can calculate the intensity of the magnetic field
in the corona as a function of $R$ and then 
compute the amount of energy flux that is
converted locally into hard X-rays, into soft quasi thermal emission
(including both intrinsic and reprocessed disc emission), into
self-absorbed synchrotron radiation and into jet-power.
In Figure 1 we show the solutions for a 10 $M_{\odot}$ black hole
accreting at $\dot m=0.005$ (typical low/hard state accretion rate,
thick lines) and at $\dot m=0.25$ (thin lines).

\begin{figure}[htb]
\centering 
\psfig{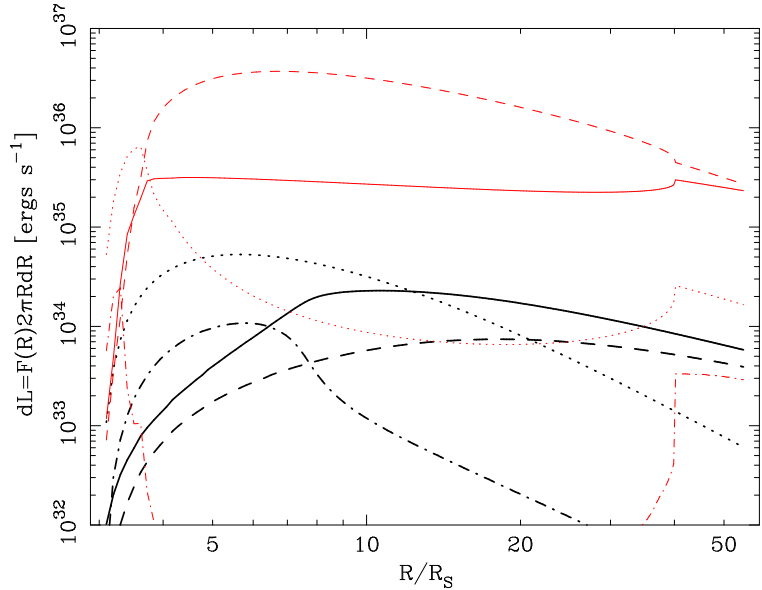}
\caption{The differential luminosity from an annulus of the accretion
  disc corona system $dL=F(R)2\pi R dR$ as a function of the distance
  $R$, for different components: hard X-rays (due to inverse Compton
  scattering in a hot corona, solid lines), soft X-rays/UV (intrinsic
  and reprocessed disc emission, dashed lines), self-absorbed
  synchrotron emission in the IR/Optical band (dot-dashed lines) and
  total kinetic jet power (dotted lines). Thick lines correspond to an
  accretion rate of $\dot m=0.005$, thin lines to $\dot m=0.25$.}
\label{fig:}
\end{figure}

At low accretion rates, the hard X-rays from the corona dominate over
the quasi-thermal disc radiation. In the innermost region of the
corona the energy density of the synchrotron radiation field dominates
over the external disc one, and coronal heating produces strong,
highly variable IR/Optical non-thermal emission. 
The kinetic power channeled into the jet
is the main repository of the gravitational energy of the accretion
flow in the inner 10 Schwarzschild radii.
At high accretion rates the inner part of the disc is radiation
pressure dominated, the corona is strongly suppressed, and no
strong outflow is generated, as expected.  

\section{Conclusions}

If a standard, geometrically thin and optically thick accretion disc
is coupled to a corona through buoyancy of magnetic flux generated by
MRI in the disc itself, then a powerful corona is generated whenever
gas pressure dominates over radiation pressure, i.e. at low accretion
rates.
 This implies that microquasar in the low/hard state may be characterized
 by strong coronae on top of very dim, cold discs, that may be seen
 only as reprocessors. Depending on the
 detailed coronal geometry, and in particular on its vertical extent,
 low accretion rate systems may produce strong MHD driven
 jets/outflows whose total kinetic energy flux may exceed the radiated
 one.
Beside their radio emission, such outflows should manifest themselves 
through their influence on the variability properties (noise and
QPOs), for example reducing the intensity and narrowness of such
timing features. In order to understand the complex interplay between
coronal physics and dynamics and the variability properties of
microquasars, full time- and radial-dependent models are needed. As a
first illustrative step in this direction, we have shown here a
stationary model for the radial profiles of the different spectral
components emerging from a disc-corona system at various accretion rates.


\end{document}